\documentclass[aps,prl,superscriptaddress,twocolumn,bibliography]{revtex4-1}
\usepackage{stix}
\usepackage{amsfonts}
\usepackage{amstext}
\usepackage{amssymb}
\usepackage{amsmath}
\usepackage[utf8]{inputenc}
\usepackage{graphicx}
\usepackage{graphics}
\usepackage{cancel}
\usepackage{color,xcolor}

\usepackage{hyperref}
\hypersetup{colorlinks=true,citecolor=blue,linkcolor=blue,filecolor=blue,urlcolor=blue,pdfstartview=FitH,pdfpagemode=UseNone}
\usepackage{tikz}

\usepackage{mleftright,mathtools}

\DeclareMathAlphabet\mathbfcal{OMS}{cmsy}{b}{n}

\definecolor{boxcolor}{HTML}{108f64}

\bibpunct{[}{]}{,}{n}{}{}

\begin{document}

\title{Magnetic control of Weyl nodes and wave packets in three-dimensional warped semimetals}

\author{Bruno Focassio} 

\affiliation{Laboratório Nacional de Nanotecnologia, LNNano/CNPEM, 13083-100, Campinas, SP, Brazil}

\author{Gabriel R. Schleder} 

\affiliation{Laboratório Nacional de Nanotecnologia, LNNano/CNPEM, 13083-100, Campinas, SP, Brazil}

\author{Adalberto Fazzio} 

\affiliation{Ilum School of Science,CNPEM, 13083-970 Campinas, São Paulo, Brazil}

\author{Rodrigo B. Capaz} 

\affiliation{Laboratório Nacional de Nanotecnologia, LNNano/CNPEM, 13083-100, Campinas, SP, Brazil}

\affiliation{Instituto de Fisica, Universidade Federal do Rio de Janeiro, Caixa Postal 68528, Rio de Janeiro, Brazil}

\author{Pedro V. Lopes} 

\affiliation{Instituto de Fisica, Universidade Federal do Rio de Janeiro, Caixa Postal 68528, Rio de Janeiro, Brazil}

\author{Jaime Ferreira} 

\affiliation{Centro Brasileiro de Pesquisas Físicas, Rua Dr. Xavier Sigaud, Rio de Janeiro, Brazil}

\author{Carsten Enderlein} 

\affiliation{Instituto de Fisica, Universidade Federal do Rio de Janeiro, Campus Duque de Caxias, Rio de Janeiro, Brazil}

\author{Marcello B. Silva Neto} 

\affiliation{Instituto de Fisica, Universidade Federal do Rio de Janeiro, Caixa Postal 68528, Rio de Janeiro, Brazil}
\affiliation{Laboratório Nacional de Nanotecnologia, LNNano/CNPEM, 13083-100, Campinas, SP, Brazil}

\date{\today} % Leave empty to omit a date

\relpenalty=9999
\binoppenalty=9999

\begin{abstract}

We investigate the topological phase transitions driven by band warping, $\lambda$, and a transverse magnetic field, $B$, for three-dimensional Weyl semimetals. First, we use the Chern number as a mathematical tool to derive the topological $\lambda\times B$ phase diagram. Next, we associate each of the topological sectors to a given angular momentum state of a rotating wave packet. Then we show how the position of the Weyl nodes can be manipulated by a transverse external magnetic field that ultimately quenches the wave packet rotation, first partially and then completely, thus resulting in a sequence of field-induced topological phase transitions. Finally, we calculate the current-induced magnetization and the anomalous Hall conductivity of a prototypical warped Weyl material. Both observables reflect the topological transitions associated with the wave packet rotation and can help to identify the elusive 3D quantum anomalous Hall effect in three-dimensional, warped Weyl materials. 

\end{abstract}


\maketitle

{\it Introduction} $-$ Wave packets are one of the most fundamental objects in quantum mechanics \cite{Kragh2009}. Schrödinger introduced them \cite{Schrdinger} right after the publication of his famous wave equation, inspired by the wave-particle duality proposed by de Broglie \cite{deBroglie}. One hundred years later, the de Broglie–Mackinnon wave packet has been finally observed, using paraxial space–time-coupled pulsed laser fields in the presence of anomalous group-velocity dispersion \cite{deBroglieWP}. Optical wave packets that are localized in space and time have also found applications ranging from microscopy and remote sensing, to nonlinear and quantum optics \cite{SpaceTimeWP}. Localized coherent phonon wave packets are being launched by ultrafast Coulomb forces in a scanning tunneling microscope using tip-enhanced terahertz electric fields \cite{PhononWP}. Ultrasonic acoustic wave packets have beaten the diffraction limit in the far-field for ultrasound waves \cite{UltraSonicAcousticWP}. Finally, matter wave packets can be optically manipulated for applications in ultrafast electron microscopy \cite{MatterWP}. %Having a finite Compton wavelength, wave packets can rotate about its own center of mass axis with a given orbital angular momentum and thus carry an intrinsic vorticity.

Optical vortices are paraxial vortex light beams possessing a Hilbert factor, $e^{i\ell\phi}$ \cite{OV30Years}, a phase singularity with a nonzero topological charge, $\ell$, which gives rise to a hollow intensity distribution (dark spot) and a phase front that describes a helix about the axis of propagation, ${\bf k}$ \cite{OpticalVorticesPhaseSingularity}. 
%The topological charge $\ell$ (also called the winding number) quantizes this winding such that there is a phase change of $2\pi\ell$ during a full rotation. 
This $e^{i\ell\phi}$ factor is also a characteristic feature of orbital angular momentum (similar to the azimuthal phase in the hydrogenic wave functions), and is not restricted to light beams. Phonons may acquire a nonzero angular momentum due to forces induced by the relative displacements of atoms out of their equilibrium positions \cite{PhononAngularMomentum}. Ultrasonic vortex beams have been shown experimentally to be obliquely reflected off a flat water-air interface, confirming the theoretically predicted reversals of phase rotation, topological charge, and orbital angular momentum in a reflected vortex beam \cite{AcousticVortexBeams}. Freely propagating electron beams have been also produced in laboratory, whose wavefront has a quantized topological structure arising from a singularity in phase, also taking the Hilbert form, $e^{i\ell \phi}$, with $\phi$ as the azimuthal angle about the beam axis and $\ell$ the topological charge \cite{BeamsWithAngularMomentum}. Remarkably, the angular momentum of an electron vortex beam can be manipulated by an external magnetic field and exhibits well-known magnetic manifestations such as Stern-Gerlach transport, Larmor precession, Aharonov-Bohm phases, Landau energy levels, and Zeeman splitting  \cite{ElectronVortexBeamMagField}. 

Quite recently, the universal mapping of topological singularities in ${\bf k}$-space to measurable topological observables in real space has been demonstrated \cite{UniversalMap}. The mapping is based on the spin-orbit interaction, angular momentum conservation, and the nontrivial winding of the Berry phase, and is fundamentally of topological origin. The spin-orbit interaction leads to a Berry curvature having a monopole structure, ${\mathbfcal{F}}({\bf k})={\bf k}/|{\bf k}|^3$ \cite{AngMomSpinOrbit}, which in turn, leads to the accumulation of the Berry phase about the linear momentum, ${\bf k}$, thus shifting the relative phase of the plane waves and modifying the orbital angular momentum \cite{BeamsWithAngularMomentum}. Although the mapping was demonstrated for the photonic honeycomb and Lieb lattices with vortex beams \cite{UniversalMap}, the mechanism is universal and also works for rotating wave packets in 3D Weyl systems \cite{UniversalMap}. 

In this work, we propose that the topological singularities of a nonmagnetic, 3D Weyl semimetal, with warping, $\lambda$, can be manipulated by an external magnetic field, $B$, due to its coupling to the wave packet orbital angular momentum. Weyl semimetals are ideal for this kind of manipulation, because while the robustness of Weyl nodes is ensured by topology \cite{NOGOTHEOREM}, their positions can be easily fine-tuned by external perturbations \cite{FMWSMTuningWeylNodesMagneticField,WeylSemimetalPrAlGe}. In fact, magnetic fields are being used to control the positions of the Weyl nodes and the emergence of the anomalous Hall or Nernst effects in magnetic semimetals  \cite{ModulationTopologicalElectronics,cheng2023tunable}. Here we demonstrate that, while $\lambda$ brings Weyl nodes closer together, contributing to the Berry curvature, ${\mathbfcal{F}}({\bf k})$, and to the wave packet rotation, a transverse magnetic field pushes the nodes away, quenching the rotation of the wave packet. This leads to a rich $\lambda\times B$ topological phase diagram and to fingerprints in transport that can help to identify the 3D quantum anomalous Hall effect, when a magnetization is induced.

{\it Warping and topology} $-$ Weyl semimetals are characterized by a linear dispersion, $E({\bf k})=\pm|{\bf k}|$, a band crossing at a Weyl node, ${\bf k}=0$, and a topological charge, $Q=\pm 1$. They are expected for systems where accidental band crossings occur and have indeed been observed for example in TaAs \cite{DiscoveryWeylSemimetals}. Multi-Weyl semimetals are characterized by a nonlinear dispersion, $E({\bf k})=\pm\sqrt{k_z^2+k_\perp^{2\ell}}$, a band crossing at a multi-Weyl node, $(k_\perp^\ell=0,k_z=0)$, and a topological charge, $Q=\pm \ell$. The existence of such multi-Weyl systems is guaranteed by point group symmetry \cite{MultiWeylTopSemimetals} and they have been experimentally observed in photonic crystals \cite{MultiWeylPhotonics}. Mixed, multi-Weyl semimetals can be constructed with a low-energy Hamiltonian 
\begin{equation}
{\cal H}_0=
\begin{bmatrix}
k_z & k_- - \lambda k_+^2 \\
k_+ - \lambda k_-^2 & -k_z 
\end{bmatrix},
\label{Multi-Weyl-Hamiltonian}
\end{equation}
therefore mixing up different topological sectors, $Q=\pm 1$ ($\lambda\ll \lambda_c$) and $Q=\mp 2$ ($\lambda\gg \lambda_c$), depending on the relative size of the warping parameter, $\lambda$, to the typical length scale associated to Weyl particles, the Compton wavelength, $\lambda_c$. This type of warping is a common feature in systems like monolayer graphene with Rashba spin-orbit coupling \cite{TrigWarpMonolayerGraphene}, twisted bilayer graphene \cite{TwistedBilayerGraphene}, anomalous Hall materials \cite{TrigWarpAnomalousHallMaterials}, monolayer MoS$_2$ \cite{MonolayerMoS2}, elemental Te \cite{Nakao1969,Hedgehog2023}, and in Bi$_2$Te$_3$ \cite{LandauLevelSpectroscopy}.

\begin{figure}[h]
\includegraphics[scale=0.32]{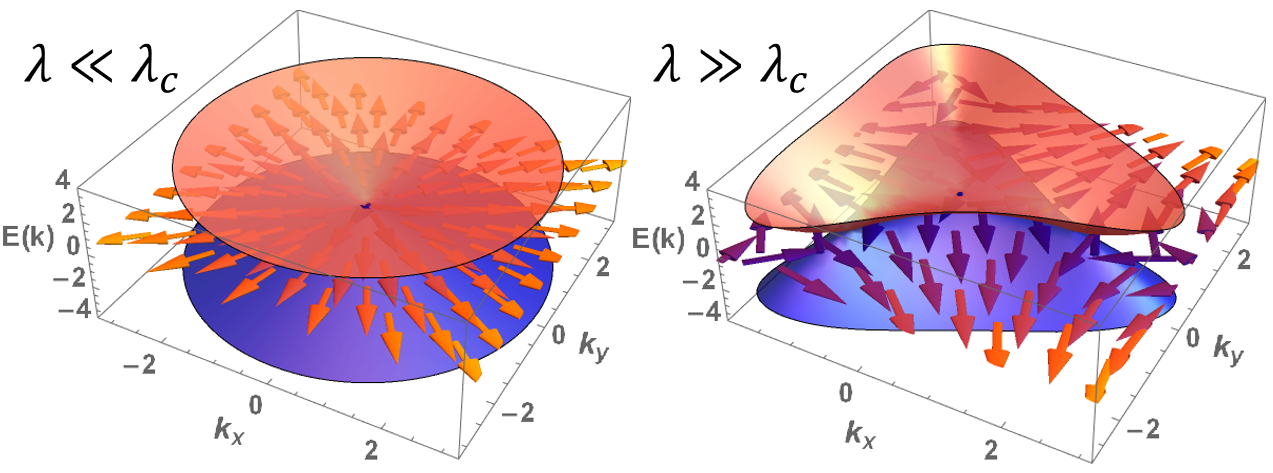}
\caption{Positive (red) and negative (blue) branches of the warped band dispersions, $E_\pm({\bf k})$, for the limiting cases, $\lambda\ll\lambda_c$ (left) and $\lambda\gg\lambda_c$ (right), and the associated orbital magnetic textures at $k_z=0$.}
\label{fig:warped.bands}
\end{figure}

Hamiltonian (\ref{Multi-Weyl-Hamiltonian}) can also be written as ${\cal H}_0={\bf d}({\bf k})\cdot{\sigma}$ with ${\bf d}({\bf k})=(d_x({\bf k}),d_y({\bf k}),d_z({\bf k}))=(k_x - \lambda (k_x^2-k_y^2),k_y + 2\lambda k_x k_y,k_z)$ and the dispersing bands, $E_\pm({\bf k})=\pm |{\bf d}({\bf k})|$, are shown in Fig. \ref{fig:warped.bands}. The eigenstates of the Hamiltonian (\ref{Multi-Weyl-Hamiltonian}) are
\begin{equation}
|u^\pm_{\bf k}\rangle=
\frac{1}{\sqrt{2|{\bf d}({\bf k})|(|{\bf d}({\bf k})|\mp d_z({\bf k}))}}
\begin{bmatrix}
d_\mp({\bf k}) \\
|{\bf d}({\bf k})|\pm d_z({\bf k})
\end{bmatrix},
\label{eq:eigenstates}
\end{equation}
where $d_\pm({\bf k})=d_x({\bf k})\pm i d_y({\bf k})$, which allows us to calculate the magnetic texture, ${\bf m}_\pm({\bf k})=\langle u^\pm_{\bf k}|\boldsymbol{\sigma}|u^\pm_{\bf k}\rangle=\pm{\bf d}({\bf k})/|{\bf d}({\bf k})|$, whose $k_z=0$ profile is also shown in Fig. \ref{fig:warped.bands}. 

%%%%%%%%%%%%%%%%%% FIGURE 1 %%%%%%%%%%%%%%%%%%%%%%%%%%
%%%%%%%%%%%%%%%%%%%%%%%%%%%%%%%%%%%%%%%%%%%%%%%%%%%%%%

\begin{figure}[h]
\includegraphics[scale=0.35]{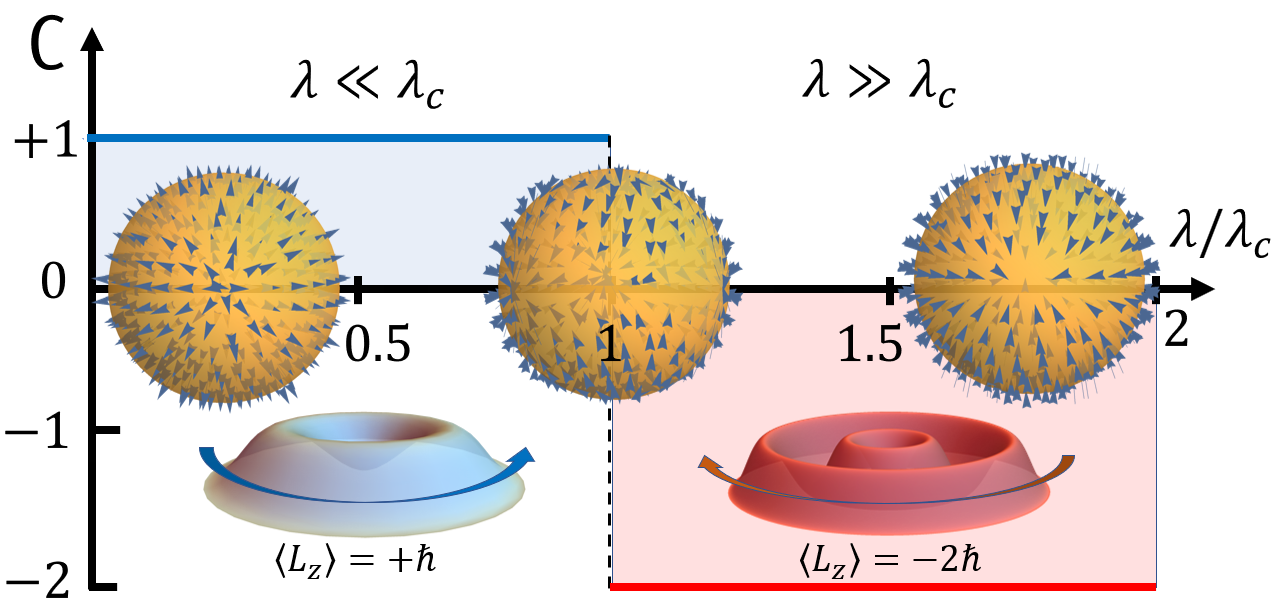}
\caption{Chern number, ${\cal C}(\lambda)$, versus warping, $\lambda$, from (\ref{AngularChern}). For $\lambda < \lambda_c$, a single Weyl node $Q=+1$ is inside the cage and ${\cal C}=+1$. The Bessel wave packet rotates counterclockwise, with $\langle L_z\rangle=+\hbar$. For $\lambda > \lambda_c$, all four nodes $Q=+1,-1,-1,-1$ are inside the cage and ${\cal C}=-2$. The Bessel wave packet rotates clockwise, with $\langle L_z\rangle=-2\hbar$.}
\label{fig:chern-vs-lambda}
\end{figure}

We will now construct a coherent wave packet following the general procedure described in \cite{CoherentWavePacket}. Since the angular momentum, $\langle L_z\rangle$, of a rotating wave packet of size $\lambda_c$ in real space is universally determined by the total topological charge, $\ell$, within an envelope of size $k_c\sim 1/\lambda_c$ in reciprocal space \cite{UniversalMap}, the Chern number becomes a very useful tool \cite{Z2Pack}. The strategy is to cage the Weyl nodes inside a closed surface in reciprocal space, for example a sphere, $S^2$, of radius $k_c\sim 1/\lambda_c$, and then to calculate the flux of the Berry curvature, $\mathbfcal{F}({\bf k})$,
\begin{equation}
    {\cal C}(\lambda)=
    \frac{1}{4\pi}\oint_{S^2} dS\;
    \hat{\bf d}({\bf k})\cdot
    \left(\nabla_{\theta}\hat{\bf d}({\bf k})\times
    \nabla_{\phi}\hat{\bf d}({\bf k})\right).
    \label{AngularChern}
\end{equation}
The Chern number, ${\cal C}(\lambda)$, defined in (\ref{AngularChern}) picks up a topological phase transition from ${\cal C}(\lambda)=+1$, for weak warping, $\lambda < \lambda_c$, to ${\cal C}(\lambda)=-2$, for strong warping, $\lambda > \lambda_c$, as shown in Fig. \ref{fig:chern-vs-lambda}.

%%%%%%%%%%%%%%%%%% FIGURE 2 %%%%%%%%%%%%%%%%%%%%%%%%%%
%%%%%%%%%%%%%%%%%%%%%%%%%%%%%%%%%%%%%%%%%%%%%%%%%%%%%%

Although the momentum-to-real-space mapping in \cite{UniversalMap} is universal, it becomes clearer in the chiral limit, $d_z=0$, when $\sigma_z{\cal H}_0\sigma_z=-{\cal H}_0$ and ${\cal H}_0$ is off-diagonal \cite{ClassificationTopInsSup3D}. By writing $d_\pm({\bf k})=|{\bf d}({\bf k})|e^{\pm i\phi_{\bf k}}$, the original eigenstates (\ref{eq:eigenstates}) simplify to
\begin{equation}
|u^\pm_{\bf k}\rangle=
\frac{1}{\sqrt{2}}
\begin{bmatrix}
e^{\mp i\phi_{\bf k}} \\
1
\end{bmatrix},
\label{eq:vortex-Weyl}
\end{equation}
with $\phi_{\bf k}=\mbox{arg}({\bf d}({\bf k}))\equiv\arctan{(d_y({\bf k})/d_x({\bf k}))}$. The ${\bf k}-$space vortex in (\ref{eq:vortex-Weyl}) is such that for any closed adiabatic path encircling the Weyl nodes, the winding of the Berry phase is \cite{UniversalMap}
\begin{equation}
    -i\oint d\phi_{\bf k}
    \langle u_{\bf k}|\nabla_{\phi_{\bf k}}|u_{\bf k}\rangle=\ell\pi.
    \label{eq:winding}
\end{equation}
Following \cite{RelativisticVortexBeams}, we now construct a wave packet with the aid of an envelope, $a({\bf k})$, of width set by the radius $k_c$ of the cage
\begin{equation}
    |W\rangle=\int \frac{d^d{\bf k}}{(2\pi)^d}\, a({\bf k})\, 
    e^{i{\bf k}\cdot({\bf r}-{\bf r}_c)}
    |u_{\bf k}\rangle,
    \label{eq:wave-packet}
\end{equation}
where ${\bf r}_c$ is the wave packet center of mass. For a paraxial wave packet, the scalar wave function acquires the Bessel form \cite{RelativisticVortexBeams}, $\psi_\ell\sim e^{i\ell\phi}J_\ell(k_\perp\rho)$, with $k_\perp=\sqrt{k_x^2+k_y^2}$ and $\rho=\sqrt{x^2+y^2}$, and a Hilbert factor, $e^{i\ell\phi}$, such that $\langle L_z\rangle=\ell\hbar$. For $\lambda < \lambda_c$, a single Weyl node $Q=+1$ is adiabatically encircled in (\ref{eq:winding}), the Chern number, ${\cal C}=+1$, generates $\ell=+1$ and the Bessel wave packet rotates counterclockwise with $\langle L_z\rangle=+\hbar$, see Fig. \ref{fig:chern-vs-lambda}. For $\lambda > \lambda_c$, however, all four Weyl nodes $Q=+1,-1,-1,-1$ are adiabatically encircled in (\ref{eq:winding}), the Chern number, ${\cal C}=-2$, generates $\ell=-2$ and the Bessel wave packet rotates clockwise with $\langle L_z\rangle=-2\hbar$, see Fig. \ref{fig:chern-vs-lambda}. 

{\it Manipulation and quenching} $-$ The semiclassical dynamics describing the motion of a rotating wave packet in the presence of electromagnetic fields was developed in \cite{Chang_2008,Chuu2010,XiaoRMP}. Uniform electric, ${\bf E}=-\nabla\phi({\bf r})-\partial_t{\bf A}({\bf r})$, and magnetic, ${\bf B}=\nabla\times{\bf A}({\bf r})$, fields can be added to the Hamiltonian (\ref{Multi-Weyl-Hamiltonian}) through both a scalar, $\phi({\bf r})$ and a vector, ${\bf A}({\bf r})$, potentials. For nonzero Berry connection, $\mathbfcal{R}$, and Berry curvature, $\mathbfcal{F}$, the dynamics of the center of mass is described by the equations of motion
\begin{eqnarray}
    \hbar \dot{\bf k}_c &=& -e{\bf E} - e\dot{\bf r}_c\times{\bf B}, \nonumber\\
    \hbar \dot{\bf r}_c &=& -i\boldsymbol{\eta}^\dagger\left[i\nabla_{{\bf k}_c}+\mathbfcal{R},{\cal H}_0\right]\boldsymbol{\eta}-\hbar \dot{\bf k}_c\times\boldsymbol{\eta}^\dagger\mathbfcal{F}\boldsymbol{\eta},
    \label{semiclassical-equations}
\end{eqnarray}
where $\dot{\bf k}_c$ and $\dot{\bf r}_c$ are the acceleration and velocity of the center of mass and $\boldsymbol{\eta}$ is the spinor component. The magnetization dynamics relative to the center of mass is described by \cite{XiaoRMP}
\begin{equation}
        i\hbar \dot{\boldsymbol{\eta}}=\left(\frac{e}{2m}\mathbfcal{L}\cdot{\bf B}-\hbar \dot{\bf k}_c\cdot\mathbfcal{R}\right)\boldsymbol{\eta},
\end{equation}
where for a relativistic particle the size of the Compton wavelenth, $\lambda_c$, and with dilation factor $\gamma$ \cite{Chang_2008,Chuu2010}
\begin{equation}
    \mathbfcal{L}({\bf k})=\frac{\hbar}{\gamma^2}\left(\sigma+\lambda_c^2\frac{{\bf k}\cdot\sigma}{\gamma+1}{\bf k}\right),
\end{equation}
is the angular momentum operator including the helicity. The full Hamiltonian describing the motion of a rotating, relativistic wave packet is, therefore, the sum of a Bloch energy, an electrostatic energy, and a Zeeman energy \cite{XiaoRMP}
\begin{equation}
    {\cal H}({\bf r}_c,{\bf k}_c)={\cal H}_0({\bf r}_c,{\bf k}_c)-e\phi({\bf r}_c)-\mathbfcal{L}({\bf k}_c)\cdot{\bf B}.
    \label{HamiltonianEMFields}
\end{equation}
In this expression, ${\cal H}_0({\bf r}_c,{\bf k}_c)$ includes all magnetic phenomena associated to the center of mass of the wave packet, ${\bf r}_c(B)$ and ${\bf k}_c(B)$, such as cyclotron orbits and Landau levels. The general theory for the magnetic susceptibility due to ${\cal H}_0({\bf r}_c,{\bf k}_c)$ is reviewed in \cite{MikitikSusceptibilityWeylNodes}. In what follows, however, we shall, instead, discuss the contribution from the rotation of the wave packet, {\it relative to the center of mass}, which corresponds to the Zeeman Hamiltonian, ${\cal H}_Z=-\mathbfcal{L}({\bf k}_c)\cdot{\bf B}$, present in (\ref{HamiltonianEMFields}).

\begin{figure}[h]
\includegraphics[scale=0.64]{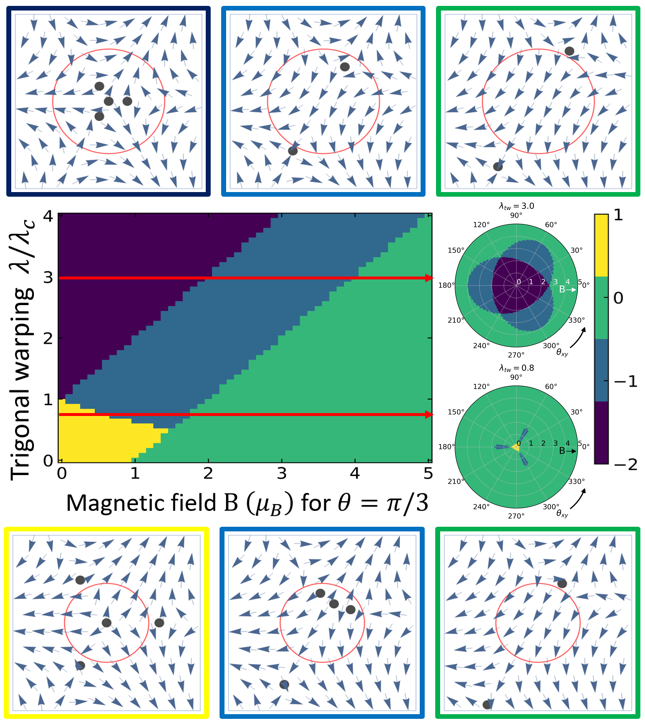}
\caption{Topological phase diagram and texture ${\bf m}({\bf k})$ for $\theta=\pi/3$. Top: partial, ${\cal C}=-2\rightarrow -1$, and complete, ${\cal C}=-1\rightarrow 0$, quenching of the wave packet. Bottom: reversal, ${\cal C}=+1\rightarrow -1$, and quenching, ${\cal C}=-1\rightarrow 0$, of the wave packet. Inset: polar diagram for $\lambda_W\equiv\lambda/\lambda_c=3.0\mbox{ (top) and }0.8 \mbox{ (bottom)}$, $B=0\rightarrow 5 (\mu_B)$, and $0\leq \phi\leq 2\pi$.}
\label{fig:two-step-topological-transition}
\end{figure}

Let us now consider the simple case of a transverse magnetic field, ${\bf B}\perp\hat{z}$ and $\phi({\bf r}_c)=0$. For Bloch electrons at long wavelengths $\gamma\approx 1$ and therefore $\mathbfcal{L}=\hbar\sigma$ \cite{Chang_2008}. In this case, the perturbation ${\cal H}_Z=-\hbar\sigma\cdot{\bf B}$ modifies the positions of all Weyl nodes \cite{TopPTInPlaneMagneticField}, so that the magnetic texture, ${\bf m}_\pm({\bf k})=\langle u^\pm_{\bf k}|\boldsymbol{\sigma}|u^\pm_{\bf k}\rangle$, becomes aligned with ${\bf B}$. Inserting ${\bf d}({\bf k},{\bf B})={\bf d}({\bf k})-{\bf B}$ into (\ref{fig:chern-vs-lambda}) we can use once again the Chern number, ${\cal C}(\lambda,B)$, to produce the $\lambda\times B$ topological phase diagram shown in Fig. \ref{fig:two-step-topological-transition}, for a magnetic field forming an angle $\theta=\pi/3$ with the $x-$axis.

%%%%%%%%%%%%%%%%%% FIGURE 3 %%%%%%%%%%%%%%%%%%%%%%%%%%
%%%%%%%%%%%%%%%%%%%%%%%%%%%%%%%%%%%%%%%%%%%%%%%%%%%%%%

At ${\bf B}=0$ and $\lambda> \lambda_c$, we have ${\cal C}(\lambda,0)=-2$. All four nodes are inside the cage, see Fig. \ref{fig:two-step-topological-transition}, and the rotating wave packet has angular momentum $\langle L_z\rangle=-2\hbar$. As the magnetic field increases, we observe, initially, the annihilation between the central node, $Q=+1$, and one of its satellites, $Q=-1$, without changing the Chern number. This occurs when the magnetic length, $l_B^{-1}=\sqrt{e B/\hbar}$, becomes comparable to the node separation \cite{pairwise-anihilation-Weyl-nodes} and the magnetic tunneling between nodes of opposite chirality is promoted \cite{WeylNodeAnihilationTaP}. Further increase in the magnetic field causes a partial quench, ${\cal C}=-2\rightarrow -1$, when one of the satellite nodes with $Q=-1$ leaves the cage, see Fig. \ref{fig:two-step-topological-transition}, and the angular momentum of the rotating wave packet becomes $\langle L_z\rangle=-\hbar$. Finally, at even higher fields the wave packet rotation is completely quenched, ${\cal C}=-1\rightarrow 0$, after all satellite nodes have left the cage and $\langle L_z\rangle=0$, see Fig. \ref{fig:two-step-topological-transition}. Variations of the angle $0\leq\theta\leq 2\pi$ produce the polar phase diagrams shown in the inset of Fig. \ref{fig:two-step-topological-transition}, which exhibits the $C_3$ symmetry of the Hamiltonian (\ref{Multi-Weyl-Hamiltonian}). Notice that for $\lambda< \lambda_c$ there exists a transition ${\cal C}=+1\rightarrow -1\rightarrow 0$, which corresponds to a single node with $Q=+1$ initially inside the cage at zero field which is joined by two of its satellites with $Q=-1$ each, see Fig. \ref{fig:two-step-topological-transition}. One of the satellites annihilates with the central one, while the other is expelled from the cage at high fields. The initial angular momentum $\langle L_z\rangle=+\hbar$ is first reversed to $\langle L_z\rangle=-\hbar$ before it is completely quenched, $\langle L_z\rangle=0$. 

%%%%%%%%%%%%%%%%%% FIGURE 4 %%%%%%%%%%%%%%%%%%%%%%%%%%
%\begin{figure}[h]
%\includegraphics[scale=0.34]{PlotPlotPhaseDiagram.png}
%\caption{Polar $\lambda\times B$ diagram for $\lambda/\lambda_c=3.0 \mbox{ (left) and } 0.8 \mbox{ (right)}$.}
%\label{fig:polarphasediagram}
%\end{figure}
%%%%%%%%%%%%%%%%%%%%%%%%%%%%%%%%%%%%%%%%%%%%%%%%%%%%%%

\begin{figure}[h]
\includegraphics[scale=0.32]{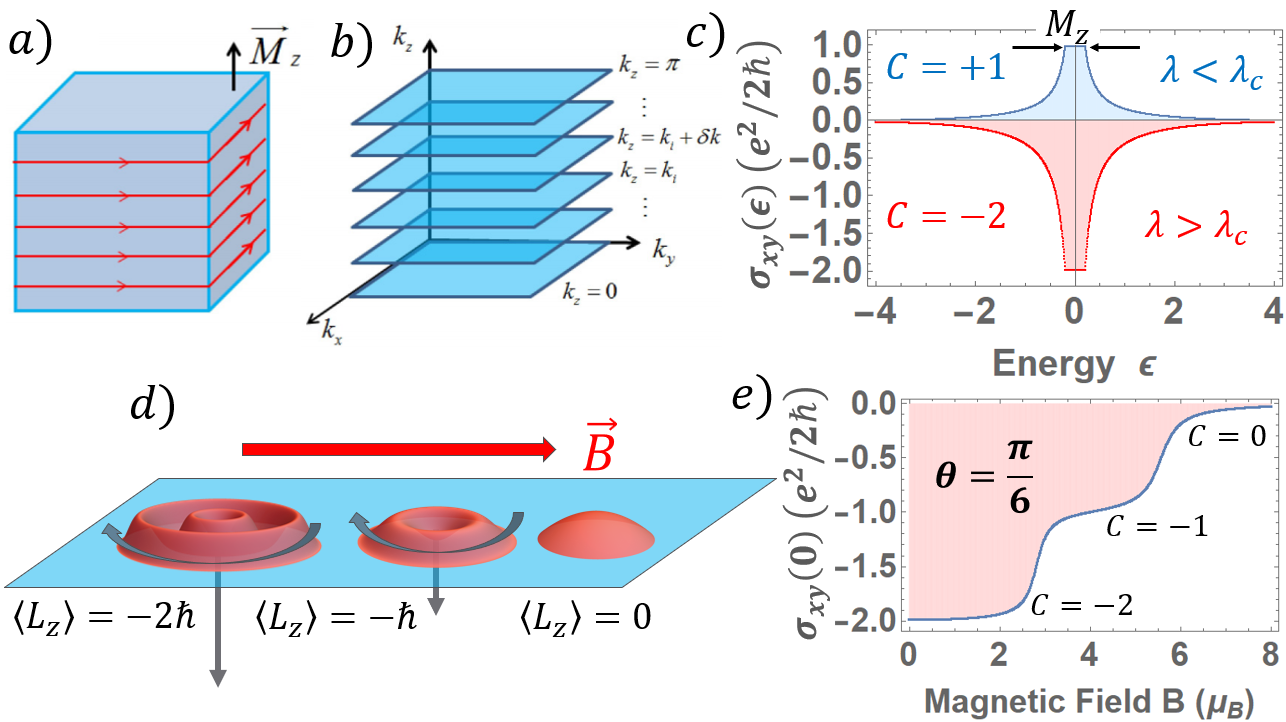}
\caption{$a)$ the current induced magnetization, $M_z({\bf E})$, and chiral surface sheet states; $b)$ adiabatically connected slices, $k_i$, perpendicular to $k_z$; $c)$ 2D Hall conductivity, $\sigma_{xy}^{2D}(\varepsilon)$, for $\lambda<\lambda_c$ (blue) and $\lambda>\lambda_c$ (red); $d)$ magnetic quenching of the wave packet for $\lambda>\lambda_c$; $e)$ 2D Hall plateaus, $\sigma_{xy}^{2D}(0,{\bf B})$ as a function of magnetic field for $\theta=\pi/6$.}
\label{fig:hall-conductivity}
\end{figure}

{\it Magnetization and the 3DQAHE} $-$ Hamiltonian (\ref{Multi-Weyl-Hamiltonian}) has recently been successfully used to describe several magneto-transport and Hall responses in trigonally warped tellurium \cite{ToyModelMagnetoTransportTellurium}. The topological transitions discussed above are bound to leave unique fingerprints in any wave packet-related properties, such as the intrinsic \cite{IntrinsicAHENonUniformElectricFieldWavePacket} or the nonlinear \cite{CovariantDerivativeBerryCurvature} anomalous Hall conductivities, $\sigma_{xy}(\varepsilon)$, or the $z-$magnetization 
\begin{equation}
    {\cal M}_z=\int\frac{d^3{\bf k}}{(2\pi)^3}f({\bf k})m_z({\bf k})+
    \frac{1}{e}\int d\varepsilon f(\varepsilon)\;\sigma_{xy}(\varepsilon),
    \label{eq:total-magnetization}
\end{equation}
where $f(\varepsilon)$ is the distribution function and $\varepsilon$ is the energy. The first term in (\ref{eq:total-magnetization}) is a smooth bulk contribution from ${\bf m}({\bf k})$ while the second term is a quantized, surface contribution from the anomalous Hall effect \cite{XiaoRMP}.
%Here the cage is substituted by the entire Brillouin zone, as indicated at the integration limit of (\ref{eq:total-magnetization}). 
In equilibrium both contributions to (\ref{eq:total-magnetization}) are identically zero by symmetry, but an electric field, ${\bf E}$, can induce a net smooth contribution to ${\cal M}_z$ \cite{FurukawaCurrIndMag,Murakami2015,SpinPolarizationHolesTellurium}, see Fig. \ref{fig:hall-conductivity}a). The center of mass distribution is governed by
\begin{equation}
    \frac{\partial f}{\partial t}+
    \dot{\bf r}_c\cdot\nabla_{{\bf r}_c}f+
    \dot{\bf k}_c\cdot\nabla_{{\bf k}_c}f=
    -\frac{f-f^{(0)}}{\tau}\equiv-\frac{\delta f}{\tau},
\end{equation}
with $\dot{\bf r}_c$ and $\dot{\bf k}_c$ defined in (\ref{semiclassical-equations}), $\tau$ is a relaxation time, and $f({\bf r}_c,{\bf k}_c,t)$ ($f^{(0)}({\bf r}_c,{\bf k}_c,t)$) is the nonequilibrium (equilibrium) distribution, such that $\delta f({\bf k},{\bf E})=-(\dot{\bf r}_c({\bf k})\tau)\cdot(e{\bf E})(\partial_\varepsilon f^{(0)}({\bf k}))$. Because of the texture-momentum locking, ${\bf m}({\bf k})$, the applied field, ${\bf E}\parallel\hat{z}$, generates an imbalance between positive and negative $k_z-$components of ${\bf m}({\bf k})$, producing a field-induced magnetization, $M_z({\bf E})\equiv\int d{\bf k}\;m_z({\bf k})\; \delta f({\bf k},{\bf E})\neq 0$ \cite{FurukawaCurrIndMag,Murakami2015,SpinPolarizationHolesTellurium}. 

%%%%%%%%%%%%%%%%%% FIGURE 5 %%%%%%%%%%%%%%%%%%%%%%%%%%
%%%%%%%%%%%%%%%%%%%%%%%%%%%%%%%%%%%%%%%%%%%%%%%%%%%%%%

\begin{figure}[h]
\includegraphics[scale=0.32]{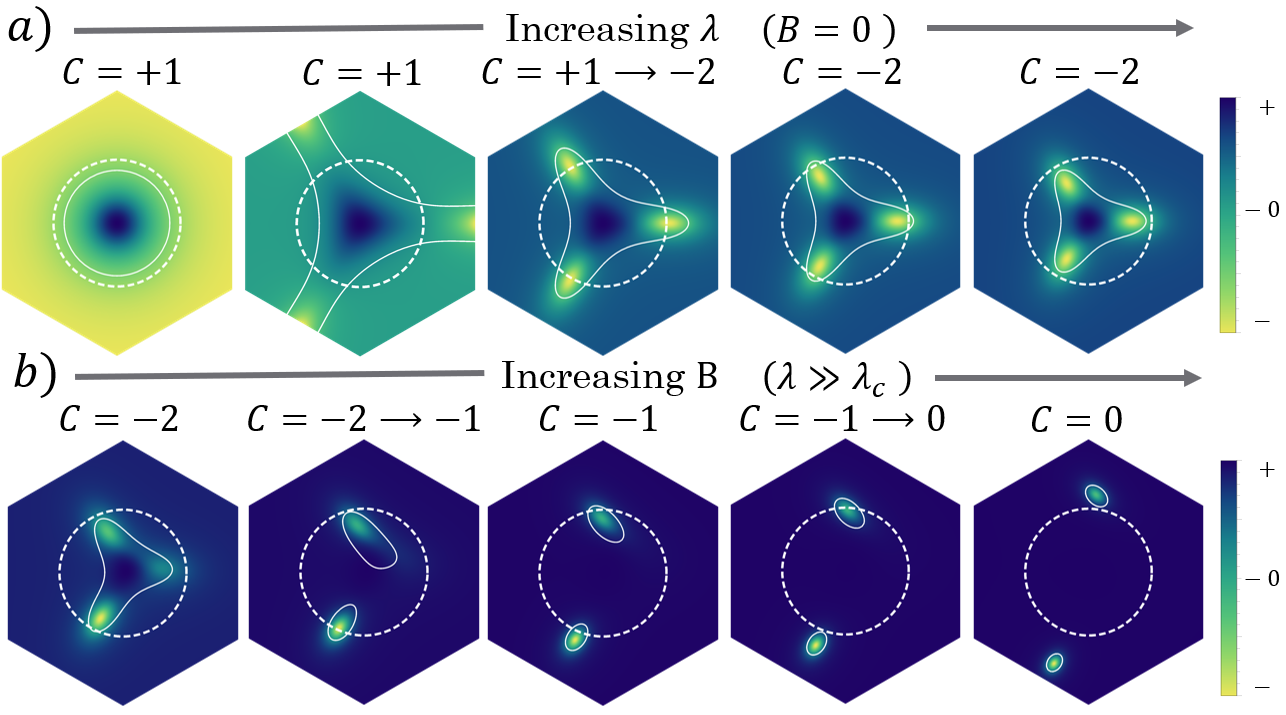}
\caption{Density plot of the Berry curvature, $\mathbfcal{F}_{xy}({\bf k},{\bf B})$, ${\bf k}-$space cages, $g({\bf k},\varepsilon,{\bf B})$ (solid white lines) with $\varepsilon>|M_z|$, and ultraviolet cutoff $|{\bf k}|<\Lambda$ (dashed white lines), inside the surroundings of a high symmetry point in the Brillouin zone. Top row: ${\bf B}=0$ and increasing $\lambda$, showing ${\cal C}=+1\rightarrow -2$; Bottom row: fixed $\lambda\gg\lambda_c$ and increasing ${\bf B}$, showing ${\cal C}=-2\rightarrow -1\rightarrow 0$, for $\theta=\pi/6$.}
\label{fig:pocket-evolution}
\end{figure}

The induced magnetization, $M_z({\bf E})$, in turn, breaks time-reversal symmetry, gaps the spectrum, and transforms the semimetal into a Chern (quantum anomalous Hall) insulator, with chiral surface sheet states, see Fig. \ref{fig:hall-conductivity}a). The Chern numbers $C_1,\dots,C_N$, associated with the slices orthogonal to the $k_z$ direction from $k_z=0$ to $k_z=\pi$, shown in Fig. \ref{fig:hall-conductivity}b), are all equal \cite{Z2Pack}. This is because planes at momenta $k_z$ and $k_z+\delta k$ can be adiabatically connected without closing the gap \cite{3DQAHEFMINS}. Hence, if any 2D cut with a given Chern number, ${\cal C}$, is identified, this material must be a 3D QAHE insulator \cite{3DQAHEFMINS}. For that reason we calculate the 2D Hall conductance at $k_z=0$ for the wave packet (\ref{eq:wave-packet}) following the procedure described in \cite{IntrinsicAHENonUniformElectricFieldWavePacket}
\begin{equation}
    \sigma_{xy}^{2D}(\varepsilon,{\bf B})=
    \frac{e^2}{2\pi\hbar}\int_{|{\bf k}|<\Lambda} \; d^2{\bf k}\; 
    \mathbfcal{F}_{xy}({\bf k},{\bf B})\; g({\bf k},\varepsilon,{\bf B}),
    \label{eq:sigmaxy2D}
\end{equation}
where $\mathbfcal{F}_{xy}({\bf k},{\bf B})=\hat{\bf d}({\bf k},{\bf B}) \cdot\left(\nabla_{k_x}\hat{\bf d}({\bf k},{\bf B})\times\nabla_{k_y}\hat{\bf d}({\bf k},{\bf B})\right)$ and $g({\bf k},\varepsilon,{\bf B})=\int d{\bf k}^\prime f(\varepsilon-|{\bf d}({\bf k}^\prime,{\bf B})|)\; |a({\bf k}-{\bf k}^\prime)|^4$, which has a compact support acting as a cage for the Berry curvature, $\mathbfcal{F}_{xy}({\bf k},{\bf B})$, and whose shape and position in ${\bf k}-$space are determined by the magnetic field, ${\bf B}$, and energy, $\varepsilon$. The ultraviolet (UV) cutoff, $\Lambda$, in (\ref{eq:sigmaxy2D}) is a material specific property that sets the scale for ${\bf k}$ above which the long-wavelength, infrared (IR) description (\ref{Multi-Weyl-Hamiltonian}) of (now) massive, $2\times2$, chiral Dirac Fermions breaks down. For $|{\bf k}|>\Lambda$, the dispersion relation bends towards the boundaries of the Brillouin zone, becoming less dispersive thus producing electronic states with a rather large effective mass at very short wavelengths, also known as {\it spectator Fermions} \cite{bernevig,HaldaneModel}. In Fig. \ref{fig:pocket-evolution}, we plot the Berry curvature, $\mathbfcal{F}_{xy}({\bf k},{\bf B})$, the cage, $g({\bf k},\varepsilon,{\bf B})$, and the UV cutoff, $\Lambda$, used in the calculation of $\sigma_{xy}^{2D}$, for different values of $\lambda$ and ${\bf B}$.

%%%%%%%%%%%%%%%%%% FIGURE 6 %%%%%%%%%%%%%%%%%%%%%%%%%%
%%%%%%%%%%%%%%%%%%%%%%%%%%%%%%%%%%%%%%%%%%%%%%%%%%%%%%

In Fig. \ref{fig:hall-conductivity}c) we show $\sigma_{xy}^{2D}(\varepsilon,{\bf B}=0)$ for $\lambda<\lambda_c$ (blue), where ${\cal C}=+1$, and $\lambda>\lambda_c$ (red), where ${\cal C}=-2$. The conductivity is nearly half-quantized, $\sigma_{xy}^{2D}(0,0)\approx{\cal C}e^2/2\hbar$, and the width of the plateaus is determined by $M_z$. For $\lambda\gg\lambda_c$ and ${\bf B}\perp\hat{z}$, the wave packet  becomes quenched, see Fig. \ref{fig:hall-conductivity}d), and the plateau of the anomalous Hall conductivity, $\sigma_{xy}^{2D}(0,{\bf B})$, crosses over markedly between $-2 e^2/2\hbar \rightarrow -1 e^2/2\hbar \rightarrow 0$, see Fig. \ref{fig:hall-conductivity}e). The absence of sharp discontinuities results from the non-singular character of the Berry curvature, $\mathbfcal{F}_{xy}({\bf k},{\bf B})$, when $M_z\neq 0$, and from the uncertainty, $\Delta{\bf k}$, brought in by the finite spread of the envelope $a({\bf k})$. The transitions occur when the cages fail to enclose regions of nonzero Berry curvature, for $|{\bf k}|<\Lambda$, see Fig. \ref{fig:pocket-evolution}. The angular momentum lost by the IR Fermions is, however, transferred to the UV {\it spectator Fermions}, so that the total topological charge inside the entire Brillouin zone is preserved, $\Delta{\cal C}_{IR}+\Delta{\cal C}_{UV}=0$, as expected.

{\it Conclusions} $-$ This work inaugurates the field of Weyl magnetronics: the magnetic control of Weyl nodes and wave packets to fine-tune the $3D$ quantum anomalous Hall effect. We also predict the existence of discretized anomalous Hall plateaus due to the wave packet rotation, even in the absense of Landau levels. Candidate Weyl magnetronic materials include monolayer and twisted bilayer graphene \cite{TrigWarpMonolayerGraphene,TwistedBilayerGraphene}, monolayer MoS$_2$ \cite{MonolayerMoS2}, and Bi$_2$Te$_3$ \cite{LandauLevelSpectroscopy}. In particular, the possibility of manipulation of Weyl nodes in elemental tellurium via strain (Weyl straintronics), numerically reported in \cite{AgapitoStrainTronics}, encourages the pursuit for the Weyl magnetronics in that very system, which might reveal the quantization of the intrinsic  \cite{IntrinsicAHENonUniformElectricFieldWavePacket} and nonlinear \cite{CovariantDerivativeBerryCurvature} anomalous Hall effects, or any other Weyl node position sensitive observables.

{\it Acknowledgements} $-$ The authors acknowledge support from CNPq - INCT (National Institute of Science and Technology on Materials Informatics), grant n. 371610/2023-0, and from FAPESP, grant. n. 2023/06522-0.

%%%%%%%%%%%%%%%%%%%%%%%%%%%%%%%%%%%%%%%%%%%%%%%%%%%%%%%%%
\bibliography{WavePacket}
%%%%%%%%%%%%%%%%%%%%%%%%%%%%%%%%%%%%%%%%%%%%%%%%%%%%%%%%%

\end{document}